\documentclass[12pt]{iopart}
\expandafter\let\csname equation*\endcsname\relax
\expandafter\let\csname endequation*\endcsname\relax

\usepackage{mathrsfs}
\usepackage{amsmath}
\usepackage{amssymb}
\usepackage{graphicx}
\usepackage{esint}
\usepackage{bm}
\usepackage{stmaryrd}
\usepackage{hyperref}
\usepackage{color}

\definecolor{myblue}{RGB}{0,0,255}
\hypersetup{
    colorlinks,
    citecolor=myblue,
    linkcolor=myblue,
    urlcolor=myblue
}

\newcommand{\mca}{\mathcal}
\newcommand{\mbb}{\mathbb}

\newcommand{\Mean}[1]{\mbb E\left[ {#1} \right]}
\newcommand{\Var}[1]{\operatorname{Var}\left[ #1\right]}
\newcommand{\MeanM}[1]{\tilde{\mbb E}\left[ {#1} \right]}
\newcommand{\VarM}[1]{\tilde{\operatorname{Var}}\left[ #1\right]}

\newcommand{\braket}[1]{\left( {#1} \right)}
\newcommand{\Braket}[1]{\left[ {#1} \right]}
\newcommand{\abs}[1]{\left| {#1} \right|}

\usepackage[whole]{bxcjkjatype}
\usepackage[normalem]{ulem}
\usepackage{soul}
\definecolor{mycyan}{RGB}{45, 99, 135}
\definecolor{mypink}{cmyk}{0, 0.7808, 0.4429, 0.1412}
\definecolor{mygreen}{RGB}{0,102,51}

\begin{document}

\title[]{Unified thermodynamic--kinetic uncertainty relation}

\author{Van Tuan Vo}
\address{Department of Information and Communication Engineering, Graduate School of Information Science and Technology, The University of Tokyo, Tokyo 113-8656, Japan}
\ead{tuan@biom.t.u-tokyo.ac.jp}

\author{Tan Van Vu}
\address{Department of Physics, Keio University, Yokohama 223-8522, Japan}
\ead{tanvu@rk.phys.keio.ac.jp}

\author{Yoshihiko Hasegawa}
\address{Department of Information and Communication Engineering, Graduate School of Information Science and Technology, The University of Tokyo, Tokyo 113-8656, Japan}
\ead{hasegawa@biom.t.u-tokyo.ac.jp}
\vspace{10pt}
\begin{abstract}
{Understanding current fluctuations is of fundamental importance and paves the way for the development of practical applications.}
According to the thermodynamic and kinetic uncertainty relations, 
the precision of currents can be constrained solely by total entropy production or dynamical activity.
In this study, we derive a tighter bound on the precision of currents in terms of both thermodynamic and kinetic quantities, demonstrating that these quantities jointly constrain current fluctuations.
The thermodynamic and kinetic uncertainty relations become particular cases of our result in asymptotic limits.
Intriguingly, the unified thermodynamic--kinetic uncertainty relation leads to a tighter classical speed limit, refining the time constraint on the system's state transformation.
The proposed framework can be extended to apply to state observables and systems with unidirectional transitions, thereby providing a constraint on the precision of the first-passage time.
\end{abstract}

\section{Introduction}
Universal relations that characterize the fluctuations of nonequilibrium systems have received considerable interest in the literature.
A prominent class of inequalities with these characteristics is the thermodynamic uncertainty relation (TUR), which was initially developed for classical systems \cite{Barato:2015:PRL, Gingrich:2016:PRL, Pietzonka:2016:PRE, Horowitz:2017:PRE, Dechant:2020:PNASUSA, Hasegawa:2019:PRE, Hasegawa:2019:PRL, VanVu:2019:PRE,  Falasco:2020:PRL, VanVu:2020:PRR, Wolpert:2020:PRL,Yoshimura:2021:PRL, Dechant:2021:PRR, Dechant:2021:PRX,  Hartich:2021:PRL, Lee:2021:PRE} (see Ref.~\cite{Horowitz:2020:NP} for a review) and subsequently extended to open quantum systems \cite{Hasegawa:2020:QTUR, Wolpert:2020:PRL, Hasegawa:2021:PRL, Miller:2021:PRL,hasegawa2021irreversibility, VanVu2021}. 
The TUR imposes an upper bound on the precision of time-integrated currents in terms of irreversible entropy production, indicating that increasing the accuracy of currents pays a price in dissipation. 
Furthermore, the kinetic uncertainty relation (KUR), which is similar but different from the TUR, imposes another upper bound on the precision of the generic counting observables in terms of the dynamical activity \cite{Garrahan2017PRE, Terlizzi2018, Hiura:2021:PRE}.
These relations are not only theoretically important but also practically relevant; they can be applied to the thermodynamic inference of dissipation even in systems with hidden degrees of freedom \cite{Li:2019:NC, Manikandan:2020:PRL, VanVu:2020:PRE, Otsubo:2020:PRE}.

The TUR and KUR have complementary roles, and there is no hierarchical relationship between them. 
For example, the TUR provides a more precise bound than the KUR when the system is close to equilibrium and the irreversible entropy production is sufficiently smaller than the dynamical activity.
In contrast, the KUR may outperform the TUR for far-from-equilibrium systems.
References \cite{Pal:2021:PhysRevResearch, Pal:2021:PhysRevResearch2} provide a perspective that the interpolated bounds of the TUR and KUR can be obtained by interpreting a bidirectional transition as two unidirectional transitions.
Nevertheless, it is important to investigate whether there exists an improved bound that incorporates both irreversible entropy production and dynamical activity to constrain the precision of currents.

Irreversible entropy production and dynamical activity are dissipative and frenetic terms, respectively, and thus represent two different facets of system dynamics \cite{MAES20201}.
Entropy production, which is the signature of nonequilibrium processes, quantifies the degree of time-reversal symmetry breaking \cite{Seifert:2012:RPP}.
In contrast, dynamical activity, quantified by the average number of jumps between system states, is time-symmetric and characterizes the timescale of systems \cite{Garrahan:2007:PRL, Maes:2008:EPL, Lecomte:2005:PRL, MAES20201}. 
Notably, these two quantities are relevant to speed limits, where they jointly constrain the speed of state transformation in both classical and quantum systems \cite{Shiraishi:2018:PRL, Gupta:2020:PRE, Yoshimura:2021:PRL, Funo:2019:NJP, VanVu:2021:PRL1, VanVu:2021:PRL2, vanvu2021finitetime, Dechant_2022, Salazar2022}.
In Ref.~\cite{Tuan:2020:PRE}, it was demonstrated that the classical speed limit and a generalized TUR have the same origin.
This close connection between the speed limits and TUR strongly suggests that both entropy production and dynamical activity may simultaneously constrain the precision of currents.

\begin{figure}
\centering
\includegraphics[width=0.9\linewidth]{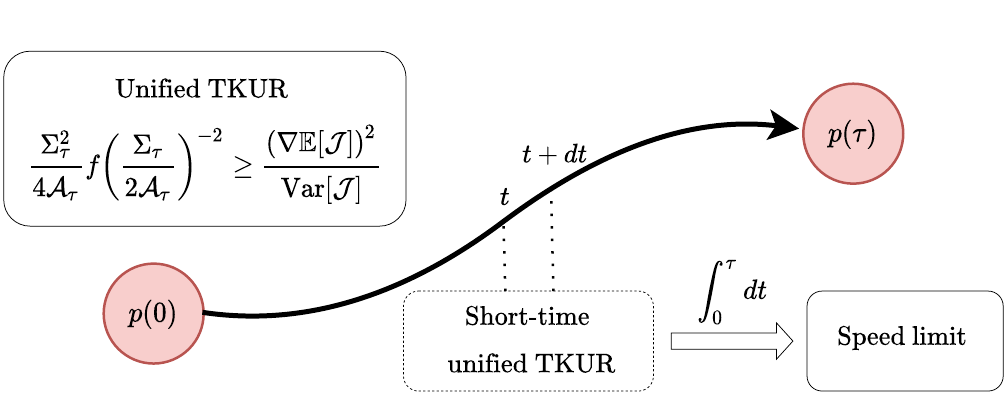} \label{FIG1}
\caption{Schematic of the relation between the unified thermodynamic--kinetic uncertainty relation (TKUR) and classical speed limit. 
According to the unified TKUR, the precision ${\braket{\nabla \Mean{\mca J}}^2} / {\Var{ \mca J}}$ of a current $\mca J$ is constrained by both the total entropy production $\Sigma_\tau$ and the dynamical activity $\mca A_\tau$. 
In the short-time limit, the precision characterizes instantaneous changes in the system distribution.
Taking the time integral of the short-time unified TKUR yields a tight classical speed limit.}
\end{figure}

In this paper, we consider discrete-state systems modeled by Markov jump processes and derive a unified bound of the thermodynamic and kinetic uncertainty relation (hereinafter referred to as the unified TKUR) for arbitrary time-integrated currents.
The upper bound of the precision consists only of entropy production and dynamical activity, indicating that the thermodynamic and kinetic costs both jointly constrain the fluctuation of currents.
Moreover, it is always stronger than the TUR and KUR, thus allowing for a more accurate TUR-based inference of dissipation.
When the ratio of entropy production to dynamical activity approaches zero or infinity, the obtained relation reduces exactly to the TUR or The unified TKUR is universal and valid for arbitrary time-dependent driving systems.
Notably, a tight classical speed limit can be obtained as a corollary of our results.
The fluctuation of currents in the short-time limit  characterizes the instantaneous change in the system distribution.
Taking the time integral of the short-time unified TKUR yields a classical speed limit, which is tighter than the bound reported in Ref.~\cite{Shiraishi:2018:PRL} (see Fig.~\ref{FIG1} for illustration).
The unified TKUR can be extended to systems with unidirectional transitions and can be applied to state observables, such as first-passage times. 

\section{Setup}
We consider a system described by a discrete-state continuous-time Markov jump process.
The system is controlled by an arbitrary protocol $\lambda_t = \lambda(v t)$ with speed parameter $v$.
The time evolution of the dynamics is governed by the master equation
\begin{align}
	\dot p_n(t, v) = \sum_m p_m(t,v)R_{nm}(\lambda_t) ,
\end{align}
where $p_n(t,v)$ is the probability of finding the system in state $n$ at time $t$ with the speed parameter $v$, and $R_{mn}(\lambda_t)$ is the transition rate from state $n$ to state $m$ controlled by the protocol $\lambda_t$.
The transition rate satisfies the normalization condition $\sum_{m} R_{mn}(\lambda_t) = 0$ and non-negativity condition $R_{mn}(\lambda_t) \geq 0$ for $n \neq m$.
We assume that the transition rates satisfy the local detailed balance condition, 
which allows us to identify the entropy flow into the environment.
In the framework of stochastic thermodynamics, the entropy production rate associated with the transition between $n$ and $m$ is defined as follows \cite{Seifert:2012:RPP}:
\begin{align}
	\sigma_{nm}(t,v) = j_{nm}(t,v) \ln \frac{p_m(t,v)R_{nm}(\lambda_t)  }{p_n(t,v)R_{mn}(\lambda_t) }.
\end{align}
where $j_{nm}(t,v) \equiv p_m(t,v)R_{nm}(\lambda_t)  - p_n(t,v)R_{mn}(\lambda_t) $ is the probability current from state $m$ to state $n$. 

Let $\omega_\tau = \{n_0, (n_1,t_1),\dots,(n_N,t_N)\}$ be a stochastic trajectory of the system during the time interval $[0, \tau]$, where the system is initially at state $n_0$ and a transition from state $n_{i-1}$ to state $n_i$ occurs at time $t_i$ for each $1\le i\le N$. 
For each trajectory, we consider a generic time-integrated current $\mca J(\omega) \equiv \sum_{i=1}^{N} d_{n_{i} n_{i-1}}$, where the increment $d_{mn}$ associated with transition $n \rightarrow m$ is anti-symmetric, $d_{mn} = - d_{nm}$.
By selecting the increments appropriately, the current will correspond to a relevant thermodynamic quantity.
For instance, $\mca J$ becomes either the stochastic entropy production current for $d_{nm}=\ln \frac{p_m(t,v)R_{nm}(\lambda_t)}{p_n(t,v)R_{mn}(\lambda_t)}$ or the heat flux into the environment for $d_{nm}=\ln \frac{R_{nm}(\lambda_t)}{R_{mn}(\lambda_t)}$.
In this time interval, the total entropy production is given by
\begin{align}
\Sigma_\tau \equiv  \int_0^\tau dt  \sum_{ n<m} \sigma_{nm}(t,v)= \int_0^\tau dt \sigma_t,
\end{align}
where the entropy production rate is defined as $\sigma_t\equiv \sum_{ n<m} \sigma_{nm}(t,v)$.
The TUR connects the total entropy production with the precision of the time-integrated current as follows \cite{Koyuk:2020:PRL}:
\begin{align}
   \frac{ \Sigma_\tau}{2} \geq \frac{\braket{\nabla \Mean{\mca J}}^2} { \Var{ \mca J}}. \label{TUR}
\end{align}
Here, $\nabla \equiv \tau \partial_\tau - v\partial_v$ is a differential operator, and $\Mean{\bullet}$ and $\Var{\bullet}$ denote the ensemble average and variance of the current, respectively.
The term $\nabla \Mean{\mca J}$ represents the response of the average current to changes in the operational time and speed of the control protocol. 
Although the TUR has a wide range of applicability, it is not tight for discrete-state systems \cite{VanVu:2020:PRE,Otsubo:2020:PRE}. 
This suggests that the total entropy production is not sufficient to characterize the current fluctuation.

To describe far-from-equilibrium systems, non-dissipative aspects must also be taken into account 
\cite{Baiesi:2009:PRL}.
One such quantity is the dynamical activity, which is the expected number of transitions between states.
The  dynamical activity is given by
\begin{align}
\mca A_\tau \equiv  \int_0^\tau dt \sum_{ n <  m} a_{nm}(t,v)= \int_0^\tau dt\, a_t,
\end{align}
where $a_{nm}(t,v) \equiv p_m(t,v)R_{nm}(\lambda_t)+ p_n(t,v)R_{mn}(\lambda_t)$ denotes the jump frequency between two states $m$ and $n$, and $a_t\equiv\sum_{ n <  m} a_{nm}(t,v)$.
For arbitrary time-dependent driven systems, the KUR implies that the precision of currents is upper bounded by the dynamical activity \cite{Koyuk:2020:PRL},
\begin{align}
    \mca A_\tau  \geq \frac{\braket{\nabla \Mean{\mca J}}^2} { \Var{ \mca J}}. \label{KUR}
\end{align}
Although we focus exclusively on generic time-integrated currents in this study, the KUR is applicable to generic counting observables.

At the transition level, we find that the probability current can be expressed in terms of the dynamical activity rate and entropy production rate (see \ref{ID}):
\begin{align}
	\frac{j_{nm}^2(t,v)}{a_{nm}(t,v)} = \frac{\sigma_{nm}^2(t,v)}{4a_{nm}(t,v)} f\braket{\frac{\sigma_{nm}(t,v)}{2 a_{nm}(t,v)} }^{-2}, \label{J_A_S}
\end{align}
where $f(x)$ is the inverse of the function $x \tanh (x)$. This provides a generic relation between the dynamical activity and entropy production. 
We define the pseudo--entropy production, which is an empirical measure of irreversibility, as follows:
\begin{align}
	 {\Sigma_\tau^{\rm ps}} \equiv  2\int_0^\tau dt \sum_{n < m}  \frac{j_{nm}(t,v) ^2}{a_{nm}(t,v)}.
\end{align}
In the overdamped Langevin limit, the pseudo--entropy production reduces to the total entropy production \cite{Shiraishi:2021:JSP}.
However, in contrast to the total entropy production, the pseudo--entropy production takes a finite value even when there exist unidirectional transitions between states. 
The pseudo--entropy production has been proven useful in deriving TURs for steady-state and periodically driven systems \cite{Shiraishi:2021:JSP, Barato:2018:NPJ, Dechant:2018:JPA}. 
Note that the function $x^2/y f({x/y})^{-2}$ is a concave function for $x, y >0$. 
Using Eq.~\eqref{J_A_S} and applying Jensen's inequality, we can calculate
\begin{equation}
\begin{aligned}[b]
\Sigma_\tau^{\rm ps}&=\int_0^\tau dt \sum_{n < m}  \frac{\sigma_{nm}^2(t,v)}{2a_{nm}(t,v)} f\braket{\frac{\sigma_{nm}(t,v)}{2 a_{nm}(t,v)} }^{-2} \\
&\le  \frac{\Braket{\int_0^\tau dt \sum_{n < m}\sigma_{nm}(t,v)}^2}{2{\int_0^\tau dt \sum_{n < m}a_{nm}(t,v)}} f\braket{\frac{\int_0^\tau dt \sum_{n < m}\sigma_{nm}(t,v)}{2\int_0^\tau dt \sum_{n < m} a_{nm}(t,v)} }^{-2},
\end{aligned}
\end{equation}
which yields an upper bound for $\Sigma_\tau^{\rm ps}$ in terms of the total entropy production and dynamical activity, as follows:
\begin{align}
	{\Sigma_\tau^{\rm ps}} \leq  \frac{ \Sigma_\tau^2}{2\mca A_\tau} f \braket{\frac{ \Sigma_\tau}{ 2\mca A_\tau}}^{-2}. \label{pseudo_S_A}
\end{align}
In the following section, we use Eq.~\eqref{pseudo_S_A} to derive a tight bound on the precision of currents for systems under time-dependent driving.

\section{Unified thermodynamic--kinetic uncertainty relation}
We consider an auxiliary system that evolves at a slightly different speed with the same control protocol.
The auxiliary density is obtained by rescaling the time $t \rightarrow  (1+\theta) t$, and modifying the speed parameter $v \rightarrow  v /(1+\theta)$ \cite{Koyuk:2020:PRL}:
\begin{align}
    \tilde{p}_{n}(t, v)  =  p_n((1+\theta)t ,  v /(1+\theta)),
\end{align}
where $\theta$ is the perturbation parameter, and the tilde represents the auxiliary dynamics.
The auxiliary dynamics is described by the following master equation:
\begin{align}
	\dot{\tilde{p}}_{n}(t, v) = \sum_m \tilde{j}_{nm} (t, v),
\end{align}
with the auxiliary probability current $\tilde{j}_{nm}(t, v) \equiv \tilde{p}_{m}(t, v)\tilde{R}_{nm}( {\lambda}_t) - \tilde{p}_{n}(t, v)\tilde{R}_{mn}({\lambda}_t)$.
According to the Cram\'er--Rao inequality, the ensemble average and variance of the time-integrated current in the auxiliary dynamics satisfy
\begin{align}
	\mca I(\theta) \geq \frac{\braket{\partial_\theta\MeanM{\mca J}}^2} {\VarM{\mca J}}, \label{CR}
\end{align}
where $\partial_\theta \equiv \frac{\partial }{\partial \theta}$ denotes the partial derivative with respect to $\theta$, and $\mca I(\theta) \equiv - \MeanM{ \partial_\theta^2 \ln \tilde{\mbb P} (\omega)}$ denotes  the Fisher information with the path probability distribution $\tilde{\mbb P} (\omega)$.
This inequality implies that the information about $\theta$ obtained by measuring the current is less than the Fisher information.

\begin{figure}
\centering
\includegraphics[width=1.0\linewidth]{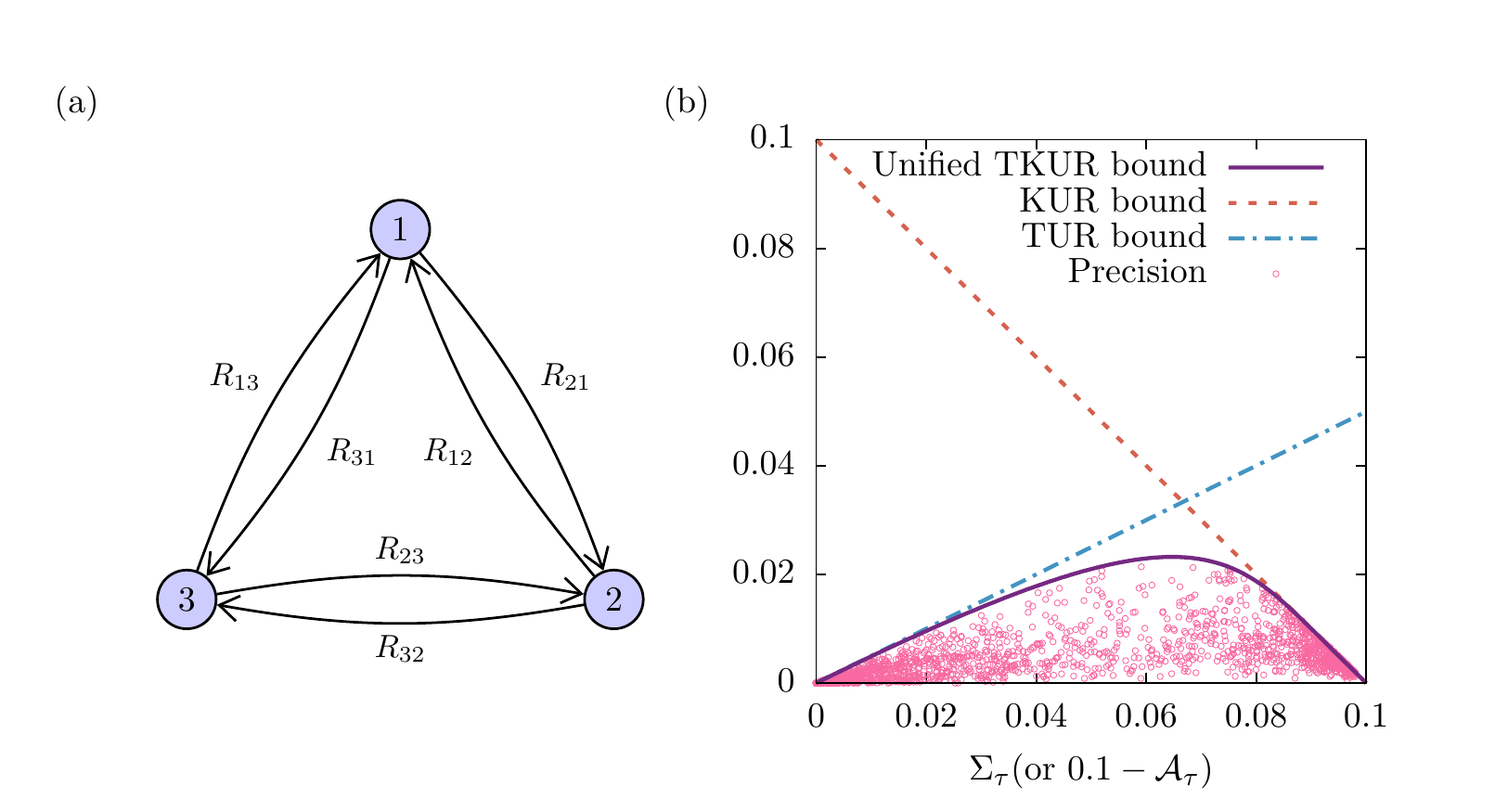}
\caption{(a) Schematic diagram of a three-state Markov jump process with fully connected states.
(b) Performance of the thermodynamic uncertainty relation (TUR), kinetic uncertainty relation (KUR) and the unified thermodynamic--kinetic uncertainty relation (TKUR) as upper bounds on the precision ${\braket{\nabla \Mean{\mca J}}^2} / {\Var{ \mca J}}$.
The transition rates between two states $m$ and $n$ are selected as $R_{mn}(\lambda_t) = 10^2 \exp\{-\mu_{mn}[f_0 + vt(f_1 -f_0)] \}$, where $\mu_{mn}$, $f_0$, and $f_1$ are randomly selected as $\mu_{mn}, f_0, f_1 \in [0, 10]$.
We consider processes in which the sum of the entropy production and dynamical activity is approximately equal to $0.1$ for time interval $\tau \in [0.1, 1]$ (i.e., $0.1-10^{-3}\le\Sigma_\tau+\mca{A}_\tau\le 0.1$).
The pink circles denote the precision where the increment $d_{mn}$ is randomly selected as $d_{mn} = -d_{nm} \in [-2, 2]$ and the speed parameter is set to $v = 1$. 
The upper bounds of Eqs.~\eqref{TUR}, \eqref{KUR}, and \eqref{main} are depicted by blue dot-dashed, orange dotted,
and violet solid lines, respectively.
}
\label{FIG2}
\end{figure}

Next, we consider the case in which the perturbation parameter $\theta$ is equal to zero.
We demonstrate that each term in Eq.~\eqref{CR} is reducible to the relevant physical quantity, under the assumption that the auxiliary probability current satisfies
\begin{align}
	\tilde{j}_{nm} (t, v)
	&= (1+\theta) j_{nm}((1+\theta)t,v/ (1+\theta)). \label{current_condition}
\end{align} 
In \ref{PSandFI}, it has been shown that the pseudo--entropy production is twice the minimum Fisher information for all possible auxiliary transition rates.
This signifies that there exists an optimal choice of the auxiliary transition rate such that 
\begin{align}
     \mca I(0) = \frac{1}{2}\Sigma_\tau^{\rm ps}.
\end{align}
It should be noted that other derivations of the TUR and KUR have also attempted to find upper bounds of $\mca I(0)$ in terms of the total entropy production of dynamical activity.
However, without considering the minimum Fisher information, the bounds are not tight, as demonstrated later.
The partial derivative of the ensemble average of the time-integrated current at $\theta = 0$ is as follows (see \ref{integrated_current}): 
\begin{align}
    \left.  \partial_\theta \MeanM{\mca J}  \right|_{\theta = 0}= \nabla\Mean{\mca J}.
\end{align}
Since the auxiliary dynamics is reduced to the original dynamics when the perturbation parameter $\theta$ is equal to zero, we have
\begin{align}
    \left.\VarM{\mca J} \right|_{\theta = 0} = \Var{\mca J}.
\end{align}
Therefore, with the optimal auxiliary transition rate, Eq.~\eqref{CR} can be rewritten as
\begin{align}
    \frac{1}{2}\Sigma_\tau^{\rm ps} \geq \frac{\braket{\nabla \Mean{\mca J}}^2} {\Var{ \mca J}}.
\end{align}
Combining with Eq.~\eqref{pseudo_S_A}, we obtain a trade-off between the precision and the thermodynamic and kinetic costs:
\begin{align}
	\mca C \equiv \frac{ \Sigma_\tau^2}{4\mca A_\tau} f \braket{\frac{ \Sigma_\tau}{ 2\mca A_\tau}}^{-2}\geq \frac{\braket{\nabla \Mean{\mca J}}^2} {\Var{ \mca J}}. \label{main}
\end{align}
We refer to Eq.~\eqref{main} as the unified TKUR, which is the central result of this paper.
Several remarks are in order.
First, the unified TKUR is always stronger than the TUR and KUR since $\mca C \leq \min\{  \Sigma_\tau / 2, \mca A_\tau \}$.
For  $ \Sigma_\tau / \mca A_\tau \gg 1$, it is reduced to the KUR, while for  $ \Sigma_\tau / \mca A_\tau \ll 1$, it reduces to the TUR. 
We numerically illustrate that the unified TKUR provides the best estimate of the precision of a random observable in a three-state Markov jump process in Fig.~\ref{FIG2}(b).
Second, since $\mca C < \min\{  \Sigma_\tau /2, \mca A_\tau \}$ when $\Sigma_\tau$ and $\mca A_\tau$ are positive and finite, our bound implies that equality of the TUR and KUR cannot be achieved in discrete-state systems, which is in agreement with the results of Refs.~\cite{VanVu:2020:PRE, Otsubo:2020:PRE}. 
In \ref{RW}, we prove that the equality condition for our relation is always satisfied for a one-dimensional biased random walk.
Third, the unified TKUR covers all use cases of the TUR, but cannot be utilized for generic counting observables like the KUR \cite{Garrahan2017PRE}. 
Finally, this relation is universal and can be extended to the cases of state observables, multidimensional control protocols, unidirectional transitions, multipartite processes, and Markovian open quantum systems.

\section{Tighter classical speed limit}
In what follows, we demonstrate that the classical speed limit is a corollary of Eq.~\eqref{main}.
We consider a system governed by a fixed control protocol in a short time interval $\Braket{t, t+\Delta t}$ when $\Delta t \rightarrow 0$. 
In this case, the unified TKUR becomes
\begin{align}
\frac{ \sigma_t^2}{4  a_t} f \braket{\frac{ \sigma_t}{ 2a_t}}^{-2}\geq \lim_{\Delta t \rightarrow0} \frac{{\Mean{\mca J(\omega_{\Delta t})}}^2} {\Delta t \Var{ \mca J (\omega_{\Delta t})}} \label{short_TUR1},
\end{align}
where $\omega_{\Delta t}$ is a trajectory in the time interval $[t, t+\Delta t]$.
We consider the integrated empirical current defined as
\begin{align}
	J_{nm}(\omega_{\Delta t}) = \sum_{i = 1 }^{N} \delta_{n_{i-1},m}\delta_{n_i, n} - \delta_{n_{i-1}, n} \delta_{n_{i},m},
\end{align}
which increases by $1$ ($-1$) when there is a jump from $m$ to $n$ ($n$ to $m$).
Here, $\delta_{x,y}$ is the Kronecker delta ($\delta_{x,y} = 1$ when $x = y$ and $0$ otherwise).
In the short-time limit, the average and variance of the integrated empirical current are as follows:
\begin{align}
	\Mean{J_{nm}(\omega_{\Delta t})}&=  \Delta t(p_m R_{nm}- p_n R_{mn} ),\\
	 \Var{J_{nm}(\omega_{\Delta t})} &= \Delta t(p_m R_{nm}+ p_n R_{mn} ),
\end{align}
respectively, to the leading order in $\Delta t$.
For simplicity, we omit the time and speed parameter notations. 
The time-integrated current is the linear combination of the empirical currents, $\mca J(\omega_{\Delta t}) =  \sum_{n < m} d_{nm} J_{nm}(\omega_{\Delta t})$.
Since all of $J_{nm}$ can be regarded as mutually independent to the leading order in $\Delta t$,
the average and variance of the time-integrated current can be written as
\begin{align}
	\Mean{\mca J(\omega_{\Delta t})}&=\sum_{n < m} d_{nm} \Mean{J_{nm}(\omega_{\Delta t})},\\
	\Var{\mca J(\omega_{\Delta t})} &=  \sum_{n < m} d_{nm}^2\Var{J_{nm}(\omega_{\Delta t})},
\end{align}
respectively.
Therefore, by setting the increment to $d_{nm} = \mathrm{sign} (p_m R_{nm}- p_n R_{mn} )$, the unified TKUR can be rewritten as follows: 
\begin{align}
\frac{ \sigma_t^2}{4  a_t} f \braket{\frac{ \sigma_t}{ 2a_t}}^{-2} \geq \frac{\Braket{\sum_{ n < m}\abs{ p_m R_{nm}- p_n R_{mn} }}^2}{a_t} .
\end{align}
Since the function $x f(x / y)^{-1}$ is a concave function for $x, y>0$, we obtain
\begin{align}
   \tau \bar{\Sigma}_\tau f \braket{\frac{\bar{\Sigma}_\tau}{2\bar{\mca A}_\tau}}^{-1} \geq 2\int_0^\tau dt \sum_{n <  m} \abs{p_m R_{nm}- p_n R_{mn} }, \label{CSL_1}
\end{align}
where $\bar{\Sigma}_\tau \equiv \tau^{-1} \Sigma_\tau$ and $\bar{\mca A}_\tau \equiv \tau^{-1} \mca A_\tau $ are the time average of the total entropy production and dynamical activity, respectively.
According to the triangle inequality and the master equation, the right-hand side of Eq.~\eqref{CSL_1} is lower bounded as follows:
\begin{align}
	 2  \int_0^\tau dt \sum_{n <  m} \abs{p_m R_{nm}- p_n R_{mn} } &\geq  \int_0^\tau dt\sum_n \abs{ \dot p_n(t)} \nonumber \\
	&\geq \mca L(\boldsymbol p(0), \boldsymbol p (\tau)), \label{CSL_2}
\end{align}
where $ \mca L(\boldsymbol p(0), \boldsymbol p (\tau)) \equiv \sum_n \abs{p_n(0) - p_n(\tau)}$ denotes the total variation distance between the initial and final distributions.
Equations \eqref{CSL_1} and \eqref{CSL_2} directly imply the following classical speed limit:
\begin{align}
	\tau \geq \frac{ \mca L(\boldsymbol p(0), \boldsymbol p (\tau))}{ \bar{\Sigma}_\tau }{{f \braket{\frac{\bar{\Sigma}_\tau }{2  \bar{\mca A}_\tau}}}} \equiv \tau_1. \label{CSL} 
\end{align}
Thus, the minimum time required for a system to change from one state to another is determined by the dynamical activity and total entropy production.
The derivation suggests that the classical speed limit is a consequence of the short-time unified TKUR.
This perspective provides not only insight into the origin of the classical speed limit but also better constraints on the operation time.
Since $f(x) \geq \operatorname{max} \{x, \sqrt{x} \}$ for $x>0$, we have lower bounds on the derived speed limit:
\begin{align}
	\tau_1 \geq \operatorname{max} \left \{\tau_2 \equiv \frac{\mca L(\boldsymbol p(0), \boldsymbol p (\tau))}{\sqrt{ 2  \bar{\mca A}_\tau\bar{\Sigma}_\tau} }, \tau_3 \equiv \frac{\mca L(\boldsymbol p(0), \boldsymbol p (\tau))}{{2} \bar{\mca A}_\tau}\right\}.
\end{align}
It is clear that Eq.~\eqref{CSL} provides a better constraint on the operation time than the speed limit $\tau \geq \tau_2$ proposed in Ref.~\cite{Shiraishi:2018:PRL}. 
When $ \bar{\Sigma}_\tau \leq {2}\bar{\mca A}_\tau$, the bound $\tau_2$ is tighter than the bound $\tau_3$. However, if the total entropy production is relatively large compared to the dynamical activity (i.e., $\bar\Sigma_\tau\gg \bar{\mca{A}}_\tau$), then $\tau_2<\tau_3$ and the operation time can be described by $\tau_3$.
In this limit, $\tau_1$ converges to $\tau_3$, and the average dynamical activity contributes significantly to determining the operation time.

\section{Unified TKUR for first-passage times}
Using the example of the first-passage time to an absorbing state $\rm X$, we illustrate that our framework can be extended to state observables and systems with unidirectional transitions. 
A transition $n \rightarrow m$ is called unidirectional when its transition rate is positive, $R_{mn} > 0$, while the rate in the reverse direction $m \rightarrow n$ is zero.
In contrast, a transition is called bidirectional when $R_{mn} > 0$ and $R_{nm} > 0$.
We assume that all transitions are bidirectional except those connected to the absorbing state $\rm X$.
Let $\rm B$ and $\rm U$ denote the set of edges with bidirectional and unidirectional transitions, respectively.
This notation allows us to decompose the dynamical activity into two components:
\begin{align}
    \mca A_\tau^{\rm B/U} &\equiv  \int_0^\tau dt  \sum_{ n<m: (n,m)\in {\rm B/U} } a_{nm}(t,v).
\end{align}
On the other hand, entropy production can only be well defined with bidirectional transitions.
In this case, the total entropy production is given by
\begin{align}
\Sigma_\tau^{\rm B} &\equiv  \int_0^\tau dt  \sum_{ n<m: (n,m)\in {\rm B} } \sigma_{nm}(t,v).
\end{align}
From Eq.~\eqref{J_A_S} and the concavity of the function $x^2/y f({x/y})^{-2}$, we obtain an upper bound on 
the pseudo--entropy production in a system with unidirectional transitions:
\begin{align}
	\Sigma_\tau^{\rm ps} \leq \mca A_\tau^{\rm U} + \frac{ (\Sigma_\tau^{\rm B})^2}{2\mca A_\tau^{\rm B}} f \braket{\frac{ \Sigma_\tau^{\rm B}}{ 2\mca A_\tau^{\rm B}}}^{-2}. \label{pseudo_S_A_Uni}
\end{align}

For a given trajectory $\omega$, the survival time before the system reaches the absorbing state $X$ is defined as $\mca T(\omega) \equiv \sum_{n\neq X} \tau_n(\omega)$, where $\tau_n(\omega)$ is the total time that the system remains in state $n$ in trajectory $\omega$.
Since $\mca T(\omega)$ is a state observable, the partial derivative of its ensemble average in the auxiliary dynamics reads (see \ref{stateOb}) 
\begin{align}
    \left.  \partial_\theta \MeanM{\mca T}  \right|_{\theta = 0}= (\nabla-1)\Mean{\mca T}.
\end{align}
According to the Cram\'er--Rao inequality, the pseudo--entropy production is then lower bounded by  
\begin{align}
	 \Sigma_\tau^{\rm ps}\geq \frac{\Braket{(\nabla -1)\Mean{\mca T}}^2} {\Var{ \mca T}}.
\end{align}
Thus, an extended version of the unified TKUR for the state observable $\mca T$ takes the form
\begin{align}
	 \mca A_\tau^{\rm U} + \frac{ (\Sigma_\tau^{\rm B})^2}{2\mca A_\tau^{\rm B}} f \braket{\frac{ \Sigma_\tau^{\rm B}}{ 2\mca A_\tau^{\rm B}}}^{-2}\geq \frac{\Braket{(\nabla -1)\Mean{\mca T}}^2} {\Var{ \mca T}}.\label{TKUR_FPT_1}
\end{align}
Note that this relation can be optimized by considering a pair of bidirectional transitions to be two unidirectional transitions \cite{Pal:2021:PhysRevResearch2}.

Next, we consider the case of relaxation where the control parameters are constant and the observation time is infinite.
In this scenario, $\mca T(\omega)$ and $\partial_\tau \Mean{\mca T}$ are the first-passage time and the survival probability, respectively.
Since the survival probability decays exponentially, $\tau\partial_\tau \Mean{\mca T}$ vanishes as $\tau \rightarrow \infty$ \cite{Gillespie1991, Pal:2021:PhysRevResearch2}. 
Once the system reaches the absorbing state, the system state remains unchanged, resulting in $\lim_{\tau \to  \infty}\mca A_{\tau}^{\rm U} = 1$, $\lim_{\tau \to \infty}\mca A_{\tau}^{\rm B} = \mca A_{\infty}^{\rm B}$ and $\lim_{\tau \to \infty}\Sigma_{\tau}^{\rm B} = \Sigma_{\infty}^{\rm B}$.
Recasting Eq.~\eqref{TKUR_FPT_1}, we obtain the unified TKUR for first-passage times:
\begin{align}
   1 + \frac{ (\Sigma_\infty^{\rm B})^2}{2\mca A_\infty^{\rm B}} f \braket{\frac{ \Sigma_\infty^{\rm B}}{ 2\mca A_\infty^{\rm B}}}^{-2}  \geq \frac{\braket{\Mean{\mca T}}^2} { \Var{ \mca T}}.
\end{align}
Thus, the precision of the first-passage time is constrained by the dynamical activity and total entropy production of the subsystem omitting the absorbing state.
Note that this bound is stricter than the thermodynamic and kinetic bounds presented in Refs.~\cite{Pal:2021:PhysRevResearch2, Hiura:2021:PRE}.

\section{Concluding perspective}
In this paper, we provide a unified perspective on the TUR, KUR, and speed limits.
We establish a tight bound on the precision of a current using the total entropy production and dynamical activity.
Our results illustrate the cooperative role of kinetic and thermodynamic contributions in constraining the precision of currents.
The unified TKUR offers a rigorous tool for thermodynamic inference in time-dependent driving systems.
This relation also provides insight that the classical speed limit can be derived from the uncertainty relation.
This is similar to the quantum case, in which one of the interpretations of Heisenberg's uncertainty principle is the quantum speed limit \cite{Tamm:1945}.
Since quantum and classical speed limits are closely related \cite{Okuyama:2017:PRL}, the results suggest that there may be a universal relation that unifies Heisenberg's uncertainty principle and the TUR.

\section*{Acknowledgments}
This work was supported by the Ministry of Education, Culture, Sports, Science, and Technology (MEXT) KAKENHI Grant No.~JP19K12153 and JP22H03659.

\appendix

\section{Derivation of Eq.~\eqref{J_A_S}} \label{ID}
For arbitrary positive real numbers $\alpha$ and $\beta$, the following equation holds:
\begin{align}
	\tanh \braket{\frac{1}{2}\ln \braket{\frac{\alpha}{\beta} }}  &= \frac{\alpha - \beta}{\alpha + \beta} .
\end{align}
Multiplying both sides by $\frac{1}{2} \ln \braket{\frac{\alpha}{\beta} }$, we obtain
\begin{align}
    	\frac{1}{2} \ln \braket{\frac{\alpha}{\beta} } \tanh \braket{\frac{1}{2} \ln \braket{\frac{\alpha}{\beta} }}  &= \frac{1}{2} \ln \braket{\frac{\alpha}{\beta} }  \frac{\alpha - \beta}{\alpha + \beta} .
\end{align}
Applying the function $f(x)$, which is the inverse of the function $ x \tanh (x)$, yields
\begin{align}
	\frac{1}{2} \ln \braket{\frac{\alpha}{\beta} }  &= f\braket{\frac{1}{2} \ln \braket{\frac{\alpha}{\beta} }  \frac{\alpha - \beta}{\alpha + \beta}},\\
	1 &= \frac{\Braket{\ln \braket{\frac{\alpha}{\beta}}}^2}{4} f\braket{  \frac{(\alpha - \beta)\ln \braket{\frac{\alpha}{\beta} } }{2(\alpha + \beta)}  }^{-2}. 
\end{align}
This leads to the following key relation:
\begin{align}
		\frac{\braket{\alpha - \beta}^2}{\alpha + \beta}  &=\frac{\Braket{(\alpha - \beta)\ln \braket{\frac{\alpha}{\beta}}}^2}{4\braket{\alpha + \beta}} f\braket{  \frac{(\alpha - \beta)\ln \braket{\frac{\alpha}{\beta} } }{2(\alpha + \beta)}  }^{-2}. \label{key}
\end{align}
Substituting $\alpha = p_{m}(t, v)R_{n m}\left(\lambda_{t}\right)$ and $\beta = p_{n}(t, v)R_{m n}\left(\lambda_{t}\right) $ into Eq.~\eqref{key}, we obtain the following identity:
\begin{align}
	\frac{j_{nm}^2(t,v)}{a_{nm}(t,v)} = \frac{\sigma_{nm}^2(t,v)}{4a_{nm}(t,v)} f\braket{\frac{\sigma_{nm}(t,v)}{2 a_{nm}(t,v)} }^{-2}.
\end{align}

\section{Pseudo--entropy production and Fisher information}\label{PSandFI}

During the time interval $[0, \tau]$, the system evolves along a trajectory $\omega_\tau = \{n_0, (n_1,t_1),\dots,(n_N,t_N)\}$, where the system is initially in state $n_0$, and a transition from state $n_{i-1}$ to state$n_i$ occurs at time $t_i$ for each $1\le i\le N$.
Let $p_{n_0}(0)$ denote the initial probability distribution of the system.
The probability density of observing the trajectory $\omega$ is given by 
\begin{align}
	\mbb P[\omega_\tau] \equiv \exp \braket{\int_0^\tau dt \sum_n \chi_n(t) R_{nn}(\lambda_t)  +  \sum_{n \neq m} \eta_{nm}(t) \ln R_{nm}(\lambda_t)} p_{n_0}(0),
\end{align}
where $\chi_n(t) \equiv \sum_{i = 1}^N \delta_{n_{i},n}\delta_{t_i, t}$,  $\eta_{nm}(t) \equiv \sum_{i = 1}^N \delta_{n_{i-1},m}\delta_{n_{i},n}\delta_{t_i, t}$. 
From the definition, the average of all possible trajectories of $\chi_n(t)$ is the probability of finding the system in state $n$ at time $t$, given by
\begin{align}
	\Mean{\chi_n(t)} &= p_n(t, v ). \label{chi}
\end{align}
Similarly, the average of $\eta_{nm}(t)$ is the total number of transitions from state $n$ to state $m$:
\begin{align}
	\Mean{\eta_{nm}(t) } &= p_m(t, v )R_{nm}(\lambda_t). \label{eta}
\end{align}

As mentioned in the main text, we consider a small perturbation of the speed of the system, 
\begin{align}
    \tilde{p}_{n}(t , v)  &=  p_n(t^\prime ,  v^\prime), \label{p_condition}\\
    \tilde \lambda_t = \lambda (t^\prime v^\prime) &= \lambda (t v) = \lambda_t,
\end{align}
where $t^\prime \equiv (1+\theta)t $ and $v^\prime \equiv v /(1+\theta)$ with perturbation control parameter $\theta$.
The auxiliary process follows a master equation,
\begin{align}
	\dot{\tilde{p}}_{n}(t, v) = \sum_m \tilde{j}_{nm} (t, v),
\end{align}
with the auxiliary probability current $\tilde{j}_{nm} (t, v) = \tilde{p}_{m}(t, v)\tilde{R}_{nm}(\lambda_t) - \tilde{p}_{n}(t, v)\tilde{R}_{mn}( \lambda_t)$. 
The auxiliary probability current is assumed to satisfy
\begin{align}
	\tilde{j}_{nm} (t, v)
	&= (1+\theta) j_{nm}(t^\prime,v^\prime) \label{j_condition}.
\end{align} 
The first derivative with respect to $\theta$ of the right-hand side of Eq.~\eqref{j_condition} is given by 
\begin{align}
   \partial_\theta \tilde{j}_{nm} (t, v)  =  &\Braket{ \tilde{R}_{nm}( \lambda_t) \partial_\theta  \tilde p_m(t, v)  -\tilde{R}_{mn}( \lambda_t) \partial_\theta \tilde p_n(t, v) } \nonumber\\&+ \braket{ \tilde p_m(t, v) \partial_\theta  \tilde{R}_{nm}( \lambda_t) - \tilde p_n(t, v)  \partial_\theta  \tilde{R}_{mn}( \lambda_t)}. \label{j_condition2}
\end{align}
Using Eq.~\eqref{p_condition}, the first derivative with respect to $\theta$ of the left-hand side of Eq.~\eqref{j_condition} can be rewritten as 
\begin{align}
   \partial_\theta \Braket{(1+\theta) j_{nm}(t^\prime,v^\prime)}  &=  (1 + \partial_\theta)  \Braket{ p_m(t^\prime ,  v^\prime)R_{nm}(\lambda_t) -  p_n(t^\prime ,  v^\prime)R_{mn}(\lambda_t) } \nonumber\\
    &= (1 + \partial_\theta)  \Braket{\tilde{p}_m(t, v)R_{nm}(\lambda_t) - \tilde{p}_n(t, v)R_{mn}(\lambda_t)}. \label{j_condition3}
\end{align}
Combining Eqs.~\eqref{j_condition}, \eqref{j_condition2}, and \eqref{j_condition3}, we obtain a requirement for the auxiliary transition rates at $\theta = 0$:
\begin{align}
	 \left. \braket{\tilde p_m(t, v) \partial_\theta  \tilde{R}_{nm}( \lambda_t) - \tilde p_n(t, v)  \partial_\theta  \tilde{R}_{mn}( \lambda_t) } \right|_{\theta = 0}=     p_m(t, v)R_{nm} (\lambda_t)   - p_n(t, v)R_{mn}( \lambda_t). \label{R_conditon}
\end{align}

Next, we calculate the Fisher information of the auxiliary trajectory probability given by
\begin{align}
	\tilde{\mbb P}[\omega_\tau] \equiv \exp \braket{\int_0^\tau dt \sum_n {\chi}_n(t) \tilde{R}_{nn}(\lambda_t)  + \sum_{n \neq m} {\eta}_{nm}(t) \ln \tilde{R}_{nm}(\lambda_t)} p_{n_0}(0).
\end{align}
Assuming that the auxiliary and original processes start with the same initial probability distribution $p_{n_0}(0)$, which is independent of $\theta$, Fisher information is given by
\begin{equation}
	\begin{aligned}
	\mca I(\theta) &= - \MeanM{ \frac{\partial^2}{\partial \theta^2} \ln \tilde{\mbb P} (\omega)}\\
	&= -\MeanM{ \int_0^\tau dt  \frac{\partial^2}{\partial \theta^2} \braket{\sum_n {\chi}_n(t) \tilde{R}_{nn}(\lambda_t)  +  \sum_{n \neq m} {\eta}_{nm}(t) \ln \tilde{R}_{nm}(\lambda_t)}} \\
	&= -\int_0^\tau dt   \braket{ \sum_n  \tilde{p}_{n}(t, v) \frac{\partial^2}{\partial \theta^2}  \tilde{R}_{nn}(\lambda_t) +  \sum_{n \neq m}  \tilde{p}_{m}(t, v) \tilde{R}_{nm}( \lambda_t) \frac{\partial^2}{\partial \theta^2}  \ln \tilde{R}_{nm}(\lambda_t)}\\
	&= \int_0^\tau dt   \sum_{n \neq m}  \braket{\tilde{p}_{m}(t, v) \frac{\partial^2}{\partial \theta^2}  \tilde{R}_{nm}(\lambda_t) -    \tilde{p}_{m}(t, v) \tilde{R}_{nm}( \lambda_t) \frac{\partial^2}{\partial \theta^2}  \ln \tilde{R}_{nm}(\lambda_t) }\\
	&= \int_0^\tau dt  \sum_{n \neq m} \tilde{p}_{m}(t, v) \tilde{R}_{nm}( \lambda_t)   \braket{\partial_\theta \ln \tilde{R}_{nm} ( \lambda_t) }^2 
\end{aligned}
\end{equation}
Here, we use Eqs.~\eqref{chi} and \eqref{eta} in the third line and the normalization condition $\sum_{m} \tilde{R}_{mn}( \lambda_t) = 0$ in the fourth line.
Therefore, the Fisher information can be written as 
\begin{align}
	\mca I(\theta)&= \int_0^\tau dt  \sum_{n < m} \Braket{\tilde{p}_{m}(t, v) \tilde{R}_{nm}( \lambda_t)   \braket{\partial_\theta \ln \tilde{R}_{nm} ( \lambda_t) }^2 + \tilde{p}_{n}(t, v) \tilde{R}_{mn}( \lambda_t)   \braket{\partial_\theta \ln \tilde{R}_{mn} ( \lambda_t) }^2} .
\end{align}
Using the inequality $\alpha x^ 2 + \beta y ^2 \geq \frac{(\alpha x-  \beta y)^2}{\alpha  + \beta} $, we obtain a lower bound on the Fisher information:
\begin{align}
	\mca I(\theta) &\geq  \int_0^\tau dt \sum_{n < m}\frac{\braket{\tilde{p}_{m}(t, v) \tilde{R}_{nm}( \lambda_t)   \partial_\theta \ln \tilde{R}_{nm}( \lambda_t) -\tilde{p}_{n}(t, v) \tilde{R}_{mn}( \lambda_t) \partial_\theta \ln \tilde{R}_{mn}( \lambda_t)}^2}
	{\tilde{p}_{m}(t, v) \tilde{R}_{nm} ( \lambda_t)+\tilde{p}_{n}(t, v)\tilde{R}_{mn}( \lambda_t)}.
\end{align}
Using the requirement for the auxiliary transition rates in Eq.~\eqref{R_conditon}, the lower bound is equal to half of the pseudo--entropy production at $\theta = 0$:
\begin{align}
	\mca I(0) &\geq  \int_0^\tau dt \sum_{n < m}\frac{\braket{p_m (t,v) R_{nm} ( \lambda_t)-p_n(t,v) R_{mn}( \lambda_t)}^2}{p_m (t,v) R_{nm} ( \lambda_t)+p_n(t,v) R_{mn}( \lambda_t)} = \frac{1}{2} \Sigma_\tau^{\rm ps}.
\end{align}
This inequality is saturated when the auxiliary transition rate $\tilde{R}_{nm}(\lambda_t)$ $(m\neq n)$ is chosen as follows:
\begin{align}
	\tilde{R}_{nm}(\lambda_t) &= R_{nm}(\lambda_t)\left( 1 + \theta \frac{p_{m}(t^\prime, v^\prime)  R_{nm}(\lambda_t)- p_{n}(t^\prime, v^\prime)  R_{mn}(\lambda_t)}{p_{m}(t^\prime, v^\prime)  R_{nm}(\lambda_t) + p_{n}(t^\prime, v^\prime)  R_{mn}(\lambda_t)} \right).
\end{align}
Therefore, the pseudo--entropy production is twice the minimum Fisher information for all possible auxiliary transition rates.

\section{Cram\'er--Rao inequality for a time-integrated current}\label{integrated_current}
The time-integrated  current has the following form:
\begin{align}
	\mca J(\omega_\tau) = \int_0^{\tau}dt  \sum_{n\neq m} d_{nm}(\lambda_t) \eta _{nm}(t),
\end{align}
where the coefficient $d_{nm}(\lambda_t)$ satisfies $ d_{nm}(\lambda_t) = - d_{mn}(\lambda_t)$.
The average of the current is given by
\begin{align}
	 \Mean{\mca J(\omega_\tau)} = \int_0^\tau dt  \sum_{n < m} d_{nm}(\lambda_t)j_{nm}(t,v).
\end{align}
Following the Cram\'er--Rao inequality, we have
\begin{align}
	\mca I(\theta) \geq \frac{\braket{\partial_\theta \MeanM{\mca J}}^2} {\VarM{\mca J}}.
\end{align}
When $\theta = 0$, the partial derivative of the average current can be calculated as
\begin{equation}
\begin{aligned}
	\left. \partial_\theta \MeanM{\mca J}   \right|_{\theta = 0} &= \tau \left. \partial_\theta \frac{1}{\tau} \int_0^\tau dt \sum_{n < m} d_{nm}(\lambda_t)\tilde{j}(t, v) 
 \right|_{\theta = 0}\\
 &= \tau \left. \partial_\theta \frac{1}{\tau} \int_0^\tau dt \sum_{n < m} d_{nm}(\lambda({t^\prime v^\prime}))(1+\theta)j_{nm}(t^\prime,v^\prime)  
 \right|_{\theta = 0}\\
	&= \tau \left. \partial_\theta \frac{1}{\tau^\prime}  \int_0^{\tau^\prime} d t^\prime \sum_{n < m} d_{nm}(\lambda({t^\prime v^\prime}))(1+\theta)j_{nm}(t^\prime,v^\prime)  
 \right|_{\theta = 0}\\
 &=  \tau \braket{1 + \tau \partial_\tau - v \partial_v}\braket{\frac{1}{\tau}\Mean{\mca J} }\\
 &= \nabla\Mean{\mca J}, \\
\end{aligned}
\end{equation}
with $\tau^\prime = (1+\theta)\tau$ and the differential operator $\nabla \equiv \tau \partial_\tau - v \partial_v$. 
Therefore, the Cram\'er--Rao inequality at $\theta = 0$ is as follows:
\begin{align}
	\mca I(0) \geq \frac{\braket{\nabla \Mean{\mca J}}^2} {\Var{ \mca J }}.
\end{align}
It should be noted that Eq.~\eqref{main} still holds for systems driven by multiple protocols $\boldsymbol{\lambda}_{t} \equiv \boldsymbol{\lambda}_{t}\left(\boldsymbol{v} \right) \equiv\left\{\lambda_{1}\left(v_{1} t\right), \ldots, \lambda_{N_{\lambda}}\left(v_{N_{\lambda}} t\right)\right\}$, with $N_{\lambda}$ speed parameter. 
In this case, the differential operator is given by $\nabla \equiv  \tau \partial_\tau-\sum_{i} \frac{v_i\partial}{\partial v_i}$.

\section{Cram\'er--Rao inequality for a time-integrated state observable}\label{stateOb}
We consider a time-integrated state observable which has the following form:
\begin{align}
	\mca K(\omega) = \int_0^{\tau}dt  \sum_n b_n(\lambda_t)\chi_n(t),
\end{align}
where $b_n$ denotes an arbitrary increment.
Following the Cram\'er--Rao inequality, we have
\begin{align}
	\mca I(\theta) \geq \frac{\braket{\partial_\theta \MeanM{\mca K}}^2} {\VarM{\mca K}}.
\end{align}
Using the identity $\MeanM{\tilde{\chi}_n(t) } = \tilde{p}_n(t,v)$, we obtain
\begin{equation}
\begin{aligned}
    \left. \partial_\theta \MeanM{\mca K}   \right|_{\theta = 0} 
    &= \tau \left. \partial_\theta \frac{1}{\tau}\int_0^\tau dt \sum_{n} b_{n}(\lambda_t)\tilde{p}_n(t, v) 
 \right|_{\theta = 0}\\
 &= \tau \left. \partial_\theta \frac{1}{\tau^\prime} \int_0^{\tau^\prime} dt^\prime  \sum_{n} b_{n}(\lambda({t^\prime v^\prime})) p_n(t^\prime, v^\prime) \right|_{\theta = 0} \\
 &= \tau \braket{\tau \partial_\tau - v \partial_v} \braket{\frac{1}{\tau} \Mean{ \mca K}}\\
 &= (\nabla -1) \Mean{ \mca K}.
\end{aligned}
\end{equation}
Thus, at $\theta = 0$, the Cram\'er--Rao inequality becomes
\begin{align}
	\mca I(0) \geq \frac{\Braket{(\nabla-1) \Mean{\mca K}}^2} {\Var{ \mca K }}.
\end{align}
Using Eq.~\eqref{pseudo_S_A}, we obtain an uncertainty relation for the time-integrated state observable:
\begin{align}
	 \frac{ \Sigma_\tau^2}{4\mca A_\tau} f \braket{\frac{ \Sigma_\tau}{ 2\mca A_\tau}}^{-2}\geq \frac{\Braket{(\nabla -1)\Mean{\mca K}}^2} {\Var{ \mca K}}.
\end{align}

\section{Biased random walk}\label{RW}
We illustrate the unified TKUR with a simple model: a one-dimensional biased random walk on $\mathbb Z$.
We assume that the walker is in position zero at $t=0$.
From the current position $x$, the walker jumps to position $x\pm 1$ at a rate $k^\pm$.
The system evolves according to the master equation:
\begin{align}
	\dot p_x(t) = k^- p_{x-1}(t) + k^+ p_{x+1}(t) - (k^- + k^+) p_x(t),
\end{align}
where $p_x(t)$ is the probability that the system is in position $x$ at time $t$.
We consider the empirical integrated current 
\begin{align}
	J(t) = n^+(t) - n^{-}(t),
\end{align}
where $n^+(t)$ and $n^-(t)$ denote the total number of transitions that $x$ has increased and decreased, respectively, by time $t$.
After time $\tau$, the moments of $J(\tau)$ are given by
\begin{align}
    \Mean{J(\tau)} &= \tau(k^+ - k^- ),\\
    \Var{J(\tau)} &= \tau(k^+ + k^-)
\end{align}
and the dynamical activity is given by 
\begin{align}
    \mca A = \tau\braket{k^+ + k^-}.
\end{align}
We assume that the transition rates satisfy the local detailed balance relation.
Then, the total entropy production is as follows:
\begin{align}
	 \sigma = \tau (k^+ - k^-)\ln\frac{k^+}{k^-}.
\end{align}
Using Eq.~\eqref{key}, it is clear that the total entropy production, dynamical activity, and precision obey the following relation:
\begin{align}
\frac{ \sigma^2}{4\mca A} f \braket{\frac{ \sigma}{ 2\mca A}}^{-2} = \frac{\Mean{J(\tau)}^2}{\Var{J(\tau)}}.
\end{align}

\section*{References}
\bibliographystyle{iopart-num}

\begin{thebibliography}{10}
\expandafter\ifx\csname url\endcsname\relax
  \def\url#1{{\tt #1}}\fi
\expandafter\ifx\csname urlprefix\endcsname\relax\def\urlprefix{URL }\fi
\providecommand{\eprint}[2][]{\url{#2}}

\bibitem{Barato:2015:PRL}
Barato A~C and Seifert U 2015 {\em Phys. Rev. Lett.\/} {\bf 114}(15) 158101
  \urlprefix\url{https://link.aps.org/doi/10.1103/PhysRevLett.114.158101}

\bibitem{Gingrich:2016:PRL}
Gingrich T~R, Horowitz J~M, Perunov N and England J~L 2016 {\em Phys. Rev.
  Lett.\/} {\bf 116}(12) 120601
  \urlprefix\url{https://link.aps.org/doi/10.1103/PhysRevLett.116.120601}

\bibitem{Pietzonka:2016:PRE}
Pietzonka P, Barato A~C and Seifert U 2016 {\em Phys. Rev. E\/} {\bf 93}(5)
  052145 \urlprefix\url{https://link.aps.org/doi/10.1103/PhysRevE.93.052145}

\bibitem{Horowitz:2017:PRE}
Horowitz J~M and Gingrich T~R 2017 {\em Phys. Rev. E\/} {\bf 96}(2) 020103(R)
  \urlprefix\url{https://link.aps.org/doi/10.1103/PhysRevE.96.020103}

\bibitem{Dechant:2020:PNASUSA}
Dechant A and ichi Sasa S 2020 {\em Proc. Natl. Acad. Sci. U.S.A.\/} {\bf 117}
  6430--6436
  \urlprefix\url{https://www.pnas.org/doi/abs/10.1073/pnas.1918386117}

\bibitem{Hasegawa:2019:PRE}
Hasegawa Y and {Van Vu} T 2019 {\em Phys. Rev. E\/} {\bf 99}(6) 062126
  \urlprefix\url{https://link.aps.org/doi/10.1103/PhysRevE.99.062126}

\bibitem{Hasegawa:2019:PRL}
Hasegawa Y and {Van Vu} T 2019 {\em Phys. Rev. Lett.\/} {\bf 123}(11) 110602
  \urlprefix\url{https://link.aps.org/doi/10.1103/PhysRevLett.123.110602}

\bibitem{VanVu:2019:PRE}
Van~Vu T and Hasegawa Y 2019 {\em Phys. Rev. E\/} {\bf 100}(1) 012134
  \urlprefix\url{https://link.aps.org/doi/10.1103/PhysRevE.100.012134}

\bibitem{Falasco:2020:PRL}
Falasco G and Esposito M 2020 {\em Phys. Rev. Lett.\/} {\bf 125}(12) 120604
  \urlprefix\url{https://link.aps.org/doi/10.1103/PhysRevLett.125.120604}

\bibitem{VanVu:2020:PRR}
Van~Vu T and Hasegawa Y 2020 {\em Phys. Rev. Research\/} {\bf 2}(1) 013060
  \urlprefix\url{https://link.aps.org/doi/10.1103/PhysRevResearch.2.013060}

\bibitem{Wolpert:2020:PRL}
Wolpert D~H 2020 {\em Phys. Rev. Lett.\/} {\bf 125}(20) 200602
  \urlprefix\url{https://link.aps.org/doi/10.1103/PhysRevLett.125.200602}

\bibitem{Yoshimura:2021:PRL}
Yoshimura K and Ito S 2021 {\em Phys. Rev. Lett.\/} {\bf 127}(16) 160601
  \urlprefix\url{https://link.aps.org/doi/10.1103/PhysRevLett.127.160601}

\bibitem{Dechant:2021:PRR}
Dechant A and Sasa S~i 2021 {\em Phys. Rev. Research\/} {\bf 3}(4) L042012
  \urlprefix\url{https://link.aps.org/doi/10.1103/PhysRevResearch.3.L042012}

\bibitem{Dechant:2021:PRX}
Dechant A and Sasa S~i 2021 {\em Phys. Rev. X\/} {\bf 11}(4) 041061
  \urlprefix\url{https://link.aps.org/doi/10.1103/PhysRevX.11.041061}

\bibitem{Hartich:2021:PRL}
Hartich D and Godec A~c~v 2021 {\em Phys. Rev. Lett.\/} {\bf 127}(8) 080601
  \urlprefix\url{https://link.aps.org/doi/10.1103/PhysRevLett.127.080601}

\bibitem{Lee:2021:PRE}
Lee J~S, Park J~M and Park H 2021 {\em Phys. Rev. E\/} {\bf 104}(5) L052102
  \urlprefix\url{https://link.aps.org/doi/10.1103/PhysRevE.104.L052102}

\bibitem{Horowitz:2020:NP}
Horowitz J~M and Gingrich T~R 2020 {\em Nat. Phys.\/} {\bf 16} 15--20 ISSN
  1745-2481 \urlprefix\url{https://doi.org/10.1038/s41567-019-0702-6}

\bibitem{Hasegawa:2020:QTUR}
Hasegawa Y 2020 {\em Phys. Rev. Lett.\/} {\bf 125}(5) 050601
  \urlprefix\url{https://link.aps.org/doi/10.1103/PhysRevLett.125.050601}

\bibitem{Hasegawa:2021:PRL}
Hasegawa Y 2021 {\em Phys. Rev. Lett.\/} {\bf 126}(1) 010602
  \urlprefix\url{https://link.aps.org/doi/10.1103/PhysRevLett.126.010602}

\bibitem{Miller:2021:PRL}
Miller H~J~D, Mohammady M~H, Perarnau-Llobet M and Guarnieri G 2021 {\em Phys.
  Rev. Lett.\/} {\bf 126}(21) 210603
  \urlprefix\url{https://link.aps.org/doi/10.1103/PhysRevLett.126.210603}

\bibitem{hasegawa2021irreversibility}
Hasegawa Y 2021 {\em Phys. Rev. Lett.\/} {\bf 127}(24) 240602
  \urlprefix\url{https://link.aps.org/doi/10.1103/PhysRevLett.127.240602}

\bibitem{VanVu2021}
Van~Vu T and Saito K 2022 {\em Phys. Rev. Lett.\/} {\bf 128}(14) 140602
  \urlprefix\url{https://link.aps.org/doi/10.1103/PhysRevLett.128.140602}

\bibitem{Garrahan2017PRE}
Garrahan J~P 2017 {\em Phys. Rev. E\/} {\bf 95}(3) 032134
  \urlprefix\url{https://link.aps.org/doi/10.1103/PhysRevE.95.032134}

\bibitem{Terlizzi2018}
{Di Terlizzi} I and Baiesi M 2018 {\em J. Phys. A: Math. Theor.\/} {\bf 52}
  02LT03 \urlprefix\url{https://doi.org/10.1088/1751-8121/aaee34}

\bibitem{Hiura:2021:PRE}
Hiura K and Sasa S~i 2021 {\em Phys. Rev. E\/} {\bf 103}(5) L050103
  \urlprefix\url{https://link.aps.org/doi/10.1103/PhysRevE.103.L050103}

\bibitem{Li:2019:NC}
Li J, Horowitz J~M, Gingrich T~R and Fakhri N 2019 {\em Nat. Commun.\/} {\bf
  10} 1666 ISSN 2041-1723
  \urlprefix\url{https://doi.org/10.1038/s41467-019-09631-x}

\bibitem{Manikandan:2020:PRL}
Manikandan S~K, Gupta D and Krishnamurthy S 2020 {\em Phys. Rev. Lett.\/} {\bf
  124}(12) 120603
  \urlprefix\url{https://link.aps.org/doi/10.1103/PhysRevLett.124.120603}

\bibitem{VanVu:2020:PRE}
Van~Vu T, Vo V~T and Hasegawa Y 2020 {\em Phys. Rev. E\/} {\bf 101}(4) 042138
  \urlprefix\url{https://link.aps.org/doi/10.1103/PhysRevE.101.042138}

\bibitem{Otsubo:2020:PRE}
Otsubo S, Ito S, Dechant A and Sagawa T 2020 {\em Phys. Rev. E\/} {\bf 101}(6)
  062106 \urlprefix\url{https://link.aps.org/doi/10.1103/PhysRevE.101.062106}

\bibitem{Pal:2021:PhysRevResearch}
Pal A, Reuveni S and Rahav S 2021 {\em Phys. Rev. Research\/} {\bf 3}(1) 013273
  \urlprefix\url{https://link.aps.org/doi/10.1103/PhysRevResearch.3.013273}

\bibitem{Pal:2021:PhysRevResearch2}
Pal A, Reuveni S and Rahav S 2021 {\em Phys. Rev. Research\/} {\bf 3}(3)
  L032034
  \urlprefix\url{https://link.aps.org/doi/10.1103/PhysRevResearch.3.L032034}

\bibitem{MAES20201}
Maes C 2020 {\em Phys. Rep.\/} {\bf 850} 1--33 ISSN 0370-1573
  \urlprefix\url{https://www.sciencedirect.com/science/article/pii/S0370157320300120}

\bibitem{Seifert:2012:RPP}
Seifert U 2012 {\em Rep. Prog. Phys.\/} {\bf 75} 126001
  \urlprefix\url{https://doi.org/10.1088/0034-4885/75/12/126001}

\bibitem{Garrahan:2007:PRL}
Garrahan J~P, Jack R~L, Lecomte V, Pitard E, van Duijvendijk K and van Wijland
  F 2007 {\em Phys. Rev. Lett.\/} {\bf 98}(19) 195702
  \urlprefix\url{https://link.aps.org/doi/10.1103/PhysRevLett.98.195702}

\bibitem{Maes:2008:EPL}
Maes C and Neto{\v{c}}n{\'{y}} K 2008 {\em Europhys Lett.\/} {\bf 82} 30003
  \urlprefix\url{https://doi.org/10.1209/0295-5075/82/30003}

\bibitem{Lecomte:2005:PRL}
Lecomte V, Appert-Rolland C and van Wijland F 2005 {\em Phys. Rev. Lett.\/}
  {\bf 95}(1) 010601
  \urlprefix\url{https://link.aps.org/doi/10.1103/PhysRevLett.95.010601}

\bibitem{Shiraishi:2018:PRL}
Shiraishi N, Funo K and Saito K 2018 {\em Phys. Rev. Lett.\/} {\bf 121}(7)
  070601
  \urlprefix\url{https://link.aps.org/doi/10.1103/PhysRevLett.121.070601}

\bibitem{Gupta:2020:PRE}
Gupta D and Busiello D~M 2020 {\em Phys. Rev. E\/} {\bf 102}(6) 062121
  \urlprefix\url{https://link.aps.org/doi/10.1103/PhysRevE.102.062121}

\bibitem{Funo:2019:NJP}
Funo K, Shiraishi N and Saito K 2019 {\em New J. Phys.\/} {\bf 21} 013006
  \urlprefix\url{https://doi.org/10.1088/1367-2630/aaf9f5}

\bibitem{VanVu:2021:PRL1}
Van~Vu T and Hasegawa Y 2021 {\em Phys. Rev. Lett.\/} {\bf 126}(1) 010601
  \urlprefix\url{https://link.aps.org/doi/10.1103/PhysRevLett.126.010601}

\bibitem{VanVu:2021:PRL2}
Van~Vu T and Hasegawa Y 2021 {\em Phys. Rev. Lett.\/} {\bf 127}(19) 190601
  \urlprefix\url{https://link.aps.org/doi/10.1103/PhysRevLett.127.190601}

\bibitem{vanvu2021finitetime}
Van~Vu T and Saito K 2022 {\em Phys. Rev. Lett.\/} {\bf 128}(1) 010602
  \urlprefix\url{https://link.aps.org/doi/10.1103/PhysRevLett.128.010602}

\bibitem{Dechant_2022}
Dechant A 2022 {\em Journal of Physics A: Mathematical and Theoretical\/} {\bf
  55} 094001 \urlprefix\url{https://doi.org/10.1088/1751-8121/ac4ac0}

\bibitem{Salazar2022}
Salazar D~S~P 2022 Lower bound for entropy production rate in stochastic
  systems far from equilibrium \urlprefix\url{https://arxiv.org/abs/2204.00875}

\bibitem{Tuan:2020:PRE}
Vo V~T, Van~Vu T and Hasegawa Y 2020 {\em Phys. Rev. E\/} {\bf 102}(6) 062132
  \urlprefix\url{https://link.aps.org/doi/10.1103/PhysRevE.102.062132}

\bibitem{Koyuk:2020:PRL}
Koyuk T and Seifert U 2020 {\em Phys. Rev. Lett.\/} {\bf 125}(26) 260604
  \urlprefix\url{https://link.aps.org/doi/10.1103/PhysRevLett.125.260604}

\bibitem{Baiesi:2009:PRL}
Baiesi M, Maes C and Wynants B 2009 {\em Phys. Rev. Lett.\/} {\bf 103}(1)
  010602
  \urlprefix\url{https://link.aps.org/doi/10.1103/PhysRevLett.103.010602}

\bibitem{Shiraishi:2021:JSP}
Shiraishi N 2021 {\em J. Stat. Phys.\/} {\bf 185} 19 ISSN 1572-9613
  \urlprefix\url{https://doi.org/10.1007/s10955-021-02829-8}

\bibitem{Barato:2018:NPJ}
Barato A~C, Chetrite R, Faggionato A and Gabrielli D 2018 {\em New J. Phys.\/}
  {\bf 20} 103023 \urlprefix\url{https://doi.org/10.1088/1367-2630/aae512}

\bibitem{Dechant:2018:JPA}
Dechant A 2018 {\em J. Phys. A\/} {\bf 52} 035001
  \urlprefix\url{https://doi.org/10.1088/1751-8121/aaf3ff}

\bibitem{Gillespie1991}
Gillespie D~T 1991 {\em Markov processes: an introduction for physical
  scientists\/} (San Diego: Elsevier) ISBN 978-0-12-283955-9

\bibitem{Tamm:1945}
Mandelstam L and Tamm I 1945 {\em J. Phys.\/} {\bf 9} 249

\bibitem{Okuyama:2017:PRL}
Okuyama M and Ohzeki M 2018 {\em Phys. Rev. Lett.\/} {\bf 120}(7) 070402
  \urlprefix\url{https://link.aps.org/doi/10.1103/PhysRevLett.120.070402}

\end{thebibliography}
\providecommand{\newblock}{}

\end{document}